\documentclass[10pt]{iopart}
\usepackage{epsfig}
\begin{document}

\title{Beware of density dependent pair potentials}

\author{A.A. Louis} \address{Department of Chemistry, Lensfield Rd,
Cambridge CB2 1EW, UK}

\begin{abstract}
Density (or state) dependent pair potentials arise naturally from
coarse-graining procedures in many areas of condensed matter science.
However, correctly using them to calculate physical properties of
interest is subtle and cannot be uncoupled from the route by which
they were derived.  Furthermore, there is usually no unique way to
coarse-grain to an effective pair potential. Even for simple systems
like liquid Argon, the pair potential that correctly reproduces the
pair structure will not generate the right virial pressure.  Ignoring
these issues in naive applications of density dependent pair
potentials can lead to an apparent dependence of thermodynamic
properties on the ensemble within which they are calculated, as well
as other inconsistencies.  These concepts are illustrated by several
pedagogical examples, including: effective pair potentials for systems
with many-body interactions, and the mapping of charged
(Debye-H\"{u}ckel) and uncharged (Asakura-Oosawa) two-component
systems onto effective one-component ones.
\end{abstract}

\pacs{61.20.Gy}

\section{Introduction}

No known materials exist in nature whose behaviour can be completely
captured by pair potentials alone.  Even a noble gas like Argon has a
finite contribution from three-body Axilrod-Teller triple-dipole
interactions\cite{Axil43}.  Thus, the pair-potentials used to describe
condensed matter systems always arise from coarse-graining procedures,
where a subset of the degrees of freedom of the full (quantum)
statistical mechanical system are integrated out.  In the
aforementioned example, integrating the three-body interactions over
angular coordinates results in effective parameters for the Lennard
Jones (LJ) potential, which will depend on state.  In metallic
systems, integrating out the free electrons leads to a configuration
independent volume term and pair potentials that depend on the global
density\cite{Ashc78}. Alternate coarse-graining procedures for metals
such as the embedded atom method\cite{Daw84,Foil85}, effective medium
theory\cite{Jaco87}, Finnis-Sinclair potentials\cite{Finn} or glue
potentials\cite{Erco86} result in a local density dependence.

Coarse-graining methods are crucial to deriving tractable statistical
mechanical treatments of soft-matters systems, where a large number of
different length and time-scales may coexist.  An increasingly popular
coarse-graining technique consists of deriving effective potentials
and exploiting their analogy to well studied simple atomic or
molecular systems to calculate phase behaviour and
correlations\cite{Bell00,Liko01,Loui01a}.  Again, these effective
interactions are often reduced to an approximate pairwise description
with parameters that depend on state.


Direct inversion from experimental structure factors are another way
to derive the parameters for effective pairwise
potentials\cite{Reat86,Robe89}. These almost always show a dependence
on state, especially for the case of soft-matter systems. This is not
surprising, of course, since one can easily imagine that the
interactions between two effective particles depends on the overall
density.  For example, changing the concentration of a micellar
solution may affect the internal structure of the micelles, which in
turn leads to a density dependence of the effective pair interaction
between the particles.

That an effective pair potential derived in one context doesn't always
perform well in another is well known, and usually categorised as a
problem of {\em transferability}.  For example, if the parameters of
an effective pair potential depend on density (but the explicit
density dependence is not known), then a parameterisation of the
potential at $\rho_1$ is not the same as the one needed at a different
density $\rho_2$ -- the potential at $\rho_1$ is not transferable to
the state point at $\rho_2$.

  What I will endeavour to show in the present paper is that there are
deeper problems associated with the use of effective pair potentials,
even when the problem of transferability appears to be solved. The
particular example studied is pair potentials that depend on the
global density $\rho$ as $v(r;\rho)$\footnote{Effective potentials
could also depend on other state variables like temperature T, but
here I mainly focus on density dependence}.  The focus is partially
pedagogical.  For that reason rather simple models are treated, with a
special emphasis on the liquid phase.  While many of these results
have already appeared in one form or another in the literature, they
are worth repeating.

The paper is organised as follows: Section~\ref{sec:2} describes the
apparent inconsistencies that arise between the virial and
compressibility routes to thermodynamics for a simple density
dependent pair potential $v(r;\rho)$.  Section~\ref{sec:3} discusses
the effective pair potentials that result from integrating out three
and higher order many-body interactions.  The effective pair potential
that correctly describes the excess internal energy is shown to be
different from the one that correctly describes the pair
structure. These points are illustrated with a specific application from
polymer solutions.  In section~\ref{sec:4} the
McMillan-Mayer\cite{McMi45} tracing out procedure is analysed for an
exactly solvable lattice version of the Asakura-Oosawa\cite{Asak58}
model.  While this procedure maps onto a useful effective
one-component picture in the semi-grand ensemble, integrating out the
smaller particles in a canonical ensemble does not lead to an
effective Hamiltonian decomposable as a sum over independent
interactions. For charged systems the canonical ensemble is the
natural choice to integrate out microscopic co and counterions.
Again, apparent ambiguities arise when the density dependent
Debye-H\"{u}ckel potential is used to derive thermodynamics.  Finally,
conclusions from these different model calculations are summarised in
section~\ref{sec:conclusions}.


\section{General thermodynamic inconsistencies from a naive
application of density dependent pair potentials}\label{sec:2}

As a first rather general example, consider a homogeneous fluid in a
volume $V$, whose $N$ particles interact with a spherically symmetric
pair potential $v(r;\rho)$, which depends on the {\em global} density
$\rho =N/V$. There are no volume, one-body, or many-body terms. No
further assumptions as to the origin of the density dependence are
made. Two established ways to calculate the equation of state (EOS)
$Z$ and other thermodynamic properties from the correlation functions
are\cite{Hans86}:

\noindent {\bf (i)} the compressibility route:
\begin{equation}\label{eq2.1}
Z_c = \frac{\beta P}{\rho} = \int_0^\rho \frac{\partial \beta
 P(\rho')}{\partial \rho'} \frac{d\rho'}{\rho} = \int_0^\rho \left[1 -
 \rho' \hat{c}(k=0;\rho')\right] \frac{d\rho'}{\rho},
\end{equation}
where $\hat{c}(k=0;\rho')$ is the zero-wavelength component of the
Fourier Transform (FT) of the direct correlation function $c(r)$,
$\beta = 1/k_B T$, and $P$ is the pressure.  This relationship follows
from simple properties of the correlation functions and their
connections to thermodynamics in the {\em grand canonical} ensemble --
it is therefore independent of the particular form of the interactions
between the particles, which need not be pairwise
additive\cite{Hans86}.

\noindent {\bf (ii)} the virial route:

\begin{equation}\label{eq2.2}
Z_{vir}^\rho=\frac{\beta P}{\rho} = 1 - \frac{2}{3} \beta \pi \rho
\int_0^\infty r^2 \left\{r \frac{\partial v(r;\rho)}{\partial r} - 3
\rho \frac{\partial v(r;\rho)}{\partial \rho} \right\} g(r) dr,
\end{equation}
where $g(r)$ is the radial distribution function.  The standard way to
derive the virial equation is directly through the {\em canonical}
partition function
\begin{equation}\label{eq2.3}  
Q(N,V,T) = \frac{\Lambda^{-3N}}{N!} \int d{\bf r}^N \exp \left\{ -\beta
\sum_{i<j} v(r_{ij};\rho) \right\}.
\end{equation}
where $\Lambda$ is the usual thermal de Broglie wavelength.
The volume derivative in 
\begin{equation}\label{eq2.4} 
\beta P = \left(\frac{\partial\log Q(N,V,T)}{\partial V} \right)_{N,T}
\end{equation}
also acts directly on the pair potential, which brings in the extra
$\partial v(r;\rho)/\partial \rho$ term in the virial
equation~(\ref{eq2.2}), a result first pointed out in 1969 by
Ascarelli and Harrison\cite{Asca69} in the context of density
dependent pair potentials used for modelling liquid metals.  This
particular form of the virial equation is only valid for pair
potentials, but the derivation of generalisations for systems with
three-body terms is straightforward.

\begin{figure}
\centerline{\epsfig{figure=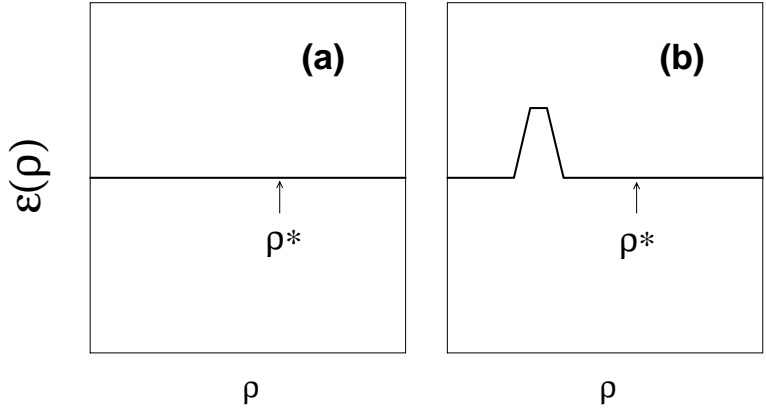,height=10cm,angle=-90}}
\caption{\label{epsrho} Amplitude $\epsilon(\rho)$ for the simple
potential $v(r;\rho)= \epsilon(\rho) v^0(r)$. Potentials (a) and (b)
are equivalent except on a small density interval.  While the
compressibility route to the EOS can distinguish between the two
potentials at $\rho*$, the virial route cannot.}
\end{figure}

So far so good: both the compressibility equation (which doesn't
change from the density independent case) and the virial equation
(which does) appear to be derived for the case of a density dependent
pair potential.  Nevertheless, this apparent rigour deceives, since it
is trivial to find density dependent pair potentials where the two
routes generate different thermodynamics.  Consider, for example, a
special class of density dependent pair potentials with
\begin{equation}\label{eq2.5}
v(r;\rho) = \epsilon(\rho) v^0(r).
\end{equation} Two possible $\epsilon(\rho)$ are shown in
Fig.~\ref{epsrho}.  The compressibility equation~(\ref{eq2.1}) results
in a different $Z$ at $\rho=\rho*$ for potentials (a) and (b), since
the effects of all densities below $\rho*$ are relevant.  In contrast,
the virial equation~(\ref{eq2.2}) cannot distinguish between the two
potentials at $\rho=\rho*$ because it only includes a {\em local}
density dependence.
  Of course it is not surprising that the two routes to thermodynamics
disagree, since one was derived in the canonical ensemble, which only
samples a single global density, while the other was derived in the
grand-canonical ensemble, which samples all densities.

However, the origin of the discrepancy lies deeper than that.  One
might think that the virial equation is less reliable, since it only
treats the density dependence as a local derivative.  But, as the
following example will show, the two routes disagree even for a simple
linear density dependence; even worse, the density dependence
correction of Ascarelli and Harrison\cite{Asca69} corrects in the
wrong direction -- it makes the discrepancy between the two routes
worse.

Consider a fluid interacting via the simple Gaussian potential
\begin{eqnarray}\label{eq2.7} 
v(r;\rho) & = &  \epsilon(\rho) \exp[-r^2] \\
\epsilon(\rho) & =& \epsilon_0 + \epsilon_1 \rho.
\end{eqnarray}
which could be viewed as a model for interactions between polymer
coils\cite{Loui00,Bolh01}. This potential falls into the class of mean
field fluids\cite{Loui00a,Lang00,Liko01b,Loui01a}, for which the
simple random phase approximation (RPA) $c(r;\rho) \approx v(r;\rho)$
is very accurate and even becomes asymptotically exact in the limit of
small $\epsilon(\rho)$ or large $\rho$.  For integrable {\em
density-independent} potentials $v(r)$, the compressibility route
leads to an EOS of the mean-field ($Z_{MF}$) form
\begin{equation}\label{eq2.8} 
Z_c^{RPA} = Z_{MF} = 1 + \frac{1}{2} \rho \beta \hat{v} (k=0),
\end{equation}
where $\hat{v}(k)$ is the FT of the potential.  For the same
integrable density independent potentials, the virial
route~(\ref{eq2.2}) reduces to
\begin{equation}\label{eq2.9}
Z_{vir}^0 = Z^{MF} - \frac{2}{3} \beta \pi \rho \int_0^\infty r^2
\left\{r \frac{\partial v(r;\rho)}{\partial r} \right\} h(r) dr,
\end{equation} where $h(r)=g(r)-1$.
  In the limit of small $\hat{v}(k=0)$ or high densities, the two
routes to $Z$ approach each other\cite{Loui00a}; the RPA closure
approximation is nearly self-consistent.

For the {\em density dependent} $v(r;\rho)$ given by
 Eq.~(\ref{eq2.7}), the compressibility equation~(\ref{eq2.1}) takes
 on a simple form:
\begin{equation}\label{eq2.10} 
Z_c^{RPA}= 1 + \pi^{3/2}\left( \frac{1}{2} \epsilon_0 \rho +
\frac{1}{3} \epsilon_1 \rho^2 \right)
\end{equation}
The virial route {\em without density-dependent corrections}, i.e.\
Eq.~(\ref{eq2.9}), results in
\begin{equation}\label{eq2.11}
Z_{vir}^0 \approx Z_{MF} = 1 + \pi^{3/2}\left( \frac{1}{2} \epsilon_0
\rho + \frac{1}{2}\epsilon_1 \rho^2\right)
\end{equation} 
for the limit of small $\hat{v}(k=0)$ or large $\rho$. Within the RPA
the corrections to Eq.~(\ref{eq2.11}) from the second term in
Eq.~(\ref{eq2.9}) can be analytically calculated for Gaussian
potentials\cite{Loui00a} and explicitly shown to be small for the
limit being considered.

\begin{figure}
\centerline{\epsfig{figure=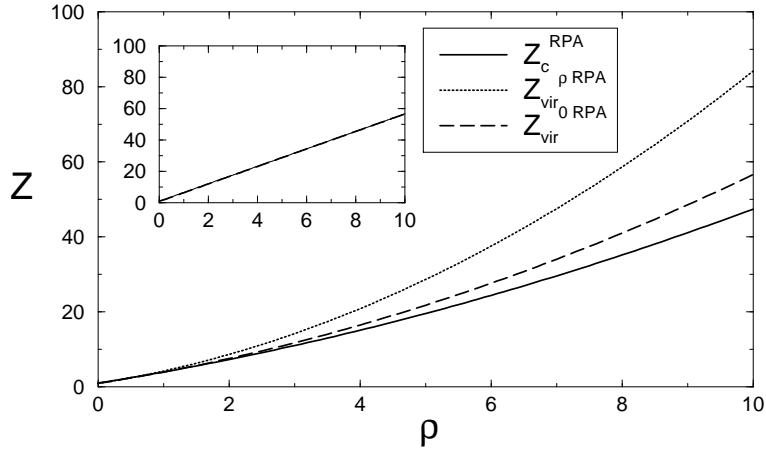,width=10cm}}
\caption{\label{Zdens} Comparison of three routes to thermodynamics
for an effective potential $\beta v(r;\rho) = (1 + 0.1\rho) \exp[-r^2]$.
$Z_c^{RPA}$ is given by Eq.~(\protect\ref{eq2.10}),
$Z_{vir}^{\rho\,\,RPA}$ comes from an analytical RPA
solution\protect\cite{unpublished} to the virial
equation~(\protect\ref{eq2.2}), and $Z_{vir}^{0\,\, RPA}$ from an
analytical RPA solution\protect\cite{Loui00a} to the simpler virial
equation~(\ref{eq2.9}).  The density-dependent correction to the
virial equation increases the disagreement between $Z_c$ and
$Z_{vir}$. Inset: In contrast to the density-dependent case, the RPA
virial and compressibility routes to $Z$ for $v(r)=2\exp[-r^2]$ agree
within the accuracy of the graph, demonstrating the near
self-consistency of the closure for such potentials.}
\end{figure}

At this level, the two routes clearly don't agree. If $\epsilon_1 > 0$
then $Z_{vir}^0 > Z_c$, while if $\epsilon_1 < 0$ then $Z_{vir}^0 <
Z_c$.  But, one might argue, the discrepancy should stem from ignoring
the density derivative term in the virial equation~(\ref{eq2.2}).
Instead, the opposite is true. Adding the density derivative
correction results in
\begin{equation}\label{eq2.12}
Z_{vir}^\rho = Z_{vir}^0 + 2 \pi \rho^2 \int r^2 g(r) \epsilon_1
\exp[-r^2]dr.
\end{equation}
Since $g(r) \geq 0$, $\epsilon_1 > 0$ implies that $Z_{vir}^\rho >
Z_{vir}^0 > Z_c$, and $\epsilon_1 < 0$ implies that $Z_{vir}^\rho <
Z_{vir}^0 < Z_c$\footnote{The conclusions for the relative ordering of
$Z_{vir}$ and $Z_c$ for $\epsilon_1 >1$ hold more generally for all
such potentials with $\partial \epsilon(\rho) /\partial \rho >1$ and
vice versa.}. In other words, the full virial equation~(\ref{eq2.2}),
derived explicitly from the canonical ensemble with a density
dependent potential, is even worse than virial
expressions~(\ref{eq2.9}) or~(\ref{eq2.11}) which ignore the density
derivative terms.  These points are illustrated in Fig.~\ref{Zdens}
for a particular example of the potential~(\ref{eq2.7}).

The problem clearly lies deeper than the fact that the virial equation
only contains a local density derivative, since, for the example
potential~(\ref{eq2.7}), this should be sufficient to describe all the
density dependence.  The next sections provide a partial answer to
this apparent conundrum by explicitly deriving density dependent
potentials, and showing that they should really be viewed as
mathematical constructs whose physical interpretation cannot be
separated from the way in which they were derived.

\section{Example 1: effective pair potentials from many-body interactions}
\label{sec:3}

This section demonstrates both in general and with an explicit example
that there is no unique way to represent averages over many-body
interactions as averages over a {\em single} pair potential that
depends on the global density $\rho$.  Although this has already been
pointed out several times in the
literature\cite{Bark69,Casa70,Rowl84,Hoef99}, mainly in the context of
noble gasses, it is worth partially repeating because it often seems
forgotten.

\subsection{Effective density dependent pair potentials from
three-body potentials}
\label{sec:3a}

Consider a homogeneous one component fluid system interacting with a
density independent pair potential $w^{(2)}({\bf r}_i,{\bf r}_j)$ and
triplet potential $w^{(3)}({\bf r}_i,{\bf r}_j,{\bf r}_k)$.  The
Hamiltonian can be written as:
\begin{equation}\label{eq3.1}
H = K + \sum_{i < j}w^{(2)}({\bf r}_j,{\bf r}_j)
 + \sum_{i <j <k}w^{(3)}({\bf r}_i,{\bf r}_j,{\bf r}_k)
\end{equation}
with $K$  the kinetic energy operator.  By using this Hamiltonian in
a standard canonical partition function, the following exact
expression can be derived for the excess internal energy $U$:
\begin{equation}\label{eq3.2}
\hspace*{-2cm} U(N,V,T) = \frac12 \rho^2 \int d{\bf r}_1 \int d{\bf r}_2 g_{12}
 w_{12} + \frac16 \rho^3 \int d{\bf r}_1 \int d{\bf r}_2 \int d{\bf r}_3
 g_{123} w_{123}
\end{equation}
where $w_{12} = w^{(2)}({\bf r}_1,{\bf r}_2)$, $w_{123} = w^{(3)}({\bf
r}_1,{\bf r}_2,{\bf r}_3)$. $g_{12}=g^{(2)}(r_{12})$ and $g_{123} =
g^{(3)}(r_{12},r_{13},r_{23})$ are the homogeneous pair and triplet
radial distribution functions respectively, and $r_{ij}$ is the
distance between particle $i$ and particle $j$

Three-body interactions are often cumbersome to use; an effective
pair-potential which reproduces the properties of the full system
would be much more convenient. A popular method to achieve this
consists of calculating the energy of the system governed by the
original many-body Hamiltonian through an accurate method (like
Eq.~(\ref{eq3.2}), and then finding an effective pair potential
$v^{eff}_U(r_{ij};\rho)$ that reproduces the same energy at the same
state-point.  For the system at hand, this can be done
explicitly\cite{Bark69,Casa70,Hoef99} by writing Eq.~(\ref{eq3.2}) as
\begin{equation}\label{eq3.2a} U(N,V,T) = \frac12 \rho^2 \int d{\bf r}_1 \int
d{\bf r}_2 g_{12}(r_{12}) v^{eff}_U(r_{12};\rho),
\end{equation} which defines the effective pair potential
\begin{equation}\label{eq3.3}
v^{eff}_U(r_{ij};\rho) = w_{12}(r_{ij}) + \delta v_U(r_{ij};\rho)
\end{equation}
where
\begin{equation}\label{eq3.4}
\delta v_U(r_{ij};\rho) = \frac13 \rho \int d{\bf r}_3 g_{13} g_{23} G_{123} w_{123}
\end{equation}
is the (additive) density dependent correction to the bare pair
potential $w_{12}(r)$.  Here $G_{123}$ is defined as the correction to
the Kirkwood superposition approximation\cite{Kirk52} for the triplet
radial distribution function:
\begin{equation}
g_{123} = g_{12}g_{13}g_{23}G_{123}.
\end{equation}

In contrast to the virial equation~(\ref{eq2.2}) of the previous
section, re-deriving the two-body energy equation for a general
$v(r;\rho)$ within the standard canonical ensemble route\cite{Hans86}
does not result in extra density dependent terms.  In other words,
within the canonical ensemble, an effective density dependent pair
interaction can be constructed that correctly captures -- by averages
over pair correlations alone -- the internal energy $U(N,V,T)$ of a
system with two and three-body interactions.  However,
$v^{eff}_U(r;\rho)$ is not sufficient for a complete pairwise
description. The three body interaction $w_{123}$ also modifies
$g_{12}$, and although $v^{eff}_U(r;\rho)$ generates a different
$g_{12}$ than would be found by using only $w_{12}$, the pair
correlations are not those of the original system, as will be shown
below.

A very useful theorem that facilitates the study of pair correlations
states that for a given homogeneous many-body system at a global
density $\rho$, there exists a bijective one to one mapping between a
unique pair potential $v^{eff}_g(r;\rho)$ and the $g^{(2)}(r)$ at that
density\cite{Hend74}\footnote{Note that this pair potential cannot
simultaneously represent the correct three-body
correlations\protect\cite{Evan90}. See e.g.\
ref.~\protect\cite{Bolh01a} for an detailed example and discussion of
this point}.  While the explicit analytical construction of
$v^{eff}_g(r;\rho)$ is not as straightforward as that of its energy
analogue $v_U^{eff}(r;\rho)$, an expansion to lowest order in density
and $w_{123}$ can be derived\cite{Casa70,Rowl84,Reat87,Atta92}:
\begin{equation}\label{eq3.5}
\hspace*{-2cm} v^{eff}_g(r_{12};\rho) \approx w_{12}(r_{12}) +
\delta v_\rho(r_{12};\rho)
= w_{12}(r_{12})  +  \rho \int d{\bf r_3}
\left\{1-\exp(-w_{123})\right\} g_{13} g_{23}.
\end{equation}
Comparing the density-dependent corrections to $v^{eff}_g(r;\rho)$ and
$v^{eff}_U(r;\rho)$ in the limit of small $\rho$ and weak $w_{123}$
leads to:
\begin{equation}\label{eq3.6}
\frac{\delta v_U(r;\rho)}{\delta v_g(r;\rho)} = \frac13 + {\cal
 O}(w_{123}^2;\rho^2).
\end{equation}
To lowest order, the two density dependent corrections differ by a
factor of three!

 The unique one-to-one correspondence between $g^{(2)}(r)$ and
$v_g^{eff}(r;\rho)$ therefore implies that $v^{eff}_U(r;\rho)$ cannot
reproduce the correct pair correlations for use in
Eq.~(\ref{eq3.2a}). This proves that the pair potential derived from
the energy equation for a many-body system cannot completely and
self-consistently capture the excess energy within a pairwise
description; $v_g^{eff}(r;\rho)$ is also needed to generate the
correct $g^{(2)}(r)$.  On the other hand, because the compressibility
equation~(\ref{eq2.1}) is independent of the underlying interactions,
$v_g^{eff}(r;\rho)$ is sufficient to derive the true compressibility,
and from it other thermodynamic quantities of interest, within a
purely pairwise description.  But the price to pay for coarse-graining
a many-body system to an effective two-body system in this way is that
the thermodynamics can only be calculated along one specific route.
Neither $v_g^{eff}(r;\rho)$ nor $v_U^{eff}(r;\rho)$ have a well
defined physical meaning independent of the way in which they were
derived.  The arguments of this subsection are not new, as the
following quote, made over 30 years ago in exactly the same context,
demonstrates:
\begin{quotation}{\em
We record our opinion that the use of density-dependent effective pair
potentials can be misleading unless it is recognised that these are
mathematical constructs to be used in specified equations rather than
physical quantities.\cite{Bark69} }
\end{quotation}

\subsection{A worked example: polymers as soft colloids}
\label{sec3:b}

The concepts of the previous section can be made more concrete by
examining a recent coarse-graining of polymers as soft
colloids\cite{Loui00,Bolh01,Bolh01a}.  This system has the advantage
that many-body correlations and interactions can be accurately
calculated from computer simulations, allowing detailed comparisons of
the coarse-grained system with the original many-body system.

\begin{figure}
\begin{center}
\centerline{\epsfig{figure=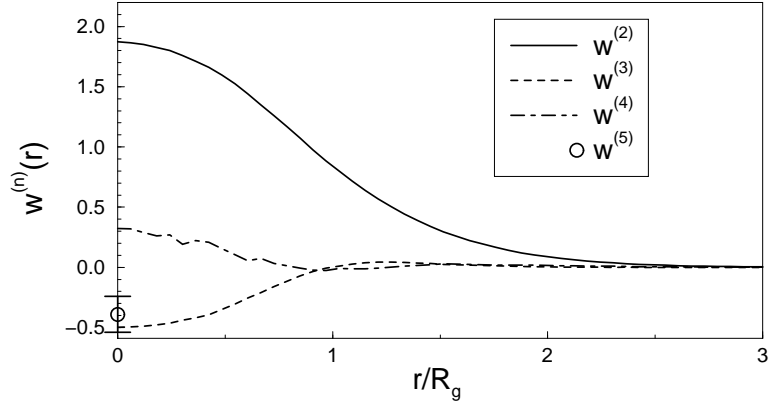,width=10cm}}
\end{center}
\caption{\label{fig:v34} Effective potentials $w^{(n)}(r)$ between the
CM of $L=500$ self avoiding walk polymer coils.  The coordinate $r$
denotes the pair distance for $n=2$, the length of an equilateral
triangle for $n=3$ and the length of a tetrahedron for $n=4$.  For
$n=5$ only the interaction at full overlap was calculated.  (taken
from Ref.~\protect\cite{Bolh01a})}
\end{figure}

The first step in the coarse-graining procedure is to choose an
effective coordinate for the polymers, which we take to be the centre
of mass (CM).  The next step is to integrate out the monomeric degrees
of freedom to derive an effective interaction between the polymer CM.
Following the discussion in\cite{Loui01a,Bolh01a}, the Helmholtz free
energy ${\cal F}$ of a set of $N$ polymers of length $L$ in a volume
$V$, with their CM distributed according to the set of coordinates
$\{{\bf r}_i\}$, can be written as the following expansion:
\begin{eqnarray}\label{eq3.10}
\hspace*{-2 cm} {\cal F}(N,V,\{{\bf r}_i\})& =& {\cal F}^{(0)}(N,V) +
\sum_{i<j}^{N} w^{(2)}({\bf r}_{i},{\bf r}_{j},) +\sum^{N}_{i<j<k }
w^{(3)}({\bf r}_{i},{\bf r}_{j},{\bf r}_{k}) + \ldots \\ \nonumber
\ldots & + & \sum^{N}_{i < \ldots < N} w^{(N)}({\bf r}_{i},{\bf
r}_{j}\ldots {\bf r}_{N})
\end{eqnarray}
 In the scaling limit, each term in the series is independent of $L$
as long as the n-tuple CM coordinates $\{ {\bf r}_{1},{\bf
r}_{2}\ldots {\bf r}_{n} \}$ are expressed in units of $R_g$, the
radius of gyration at zero density.  This coarse-grained free-energy
contains an implicit statistical average over all the monomeric
degrees of freedom for a fixed set of CM coordinates $\{{\bf r}_i\}$.
${\cal F}^{(0)}(N,V)$ is the zero-body volume term, related to the
internal free energy of a single polymer; translational symmetry
implies that there is no explicit one-body term.  The pair and higher
body terms are defined in the standard way: the $nth$ body term
$w^{(n)}({\bf r}_{1},{\bf r}_{2}\ldots {\bf r}_{i_n})$ for a
particular set of $n$ CM coordinates is given by the total free-energy
${\cal F}$ for $n$ polymers at those coordinates, minus the sum of all
the lower order terms\cite{Loui01a,Bolh01a}.  These interactions can
be explicitly calculated by computer simulations; examples are shown
in Fig.~\ref{fig:v34}, taken from ref.~\cite{Bolh01a}, where further
details can be found. The relative importance of each term decreases
for increasing $n$, so that in principle the many-body
expansion~(\ref{eq3.10}) is expected to converge\cite{Bolh01a}.

 The free energy $F(N,V)$ of the underlying polymer system follows
 from a final trace over all CM coordinates
\begin{equation}\label{eq3.11}
F(N,V) = -\log \sum_{\{{\bf r}_i\}} \exp \left[-{\cal F}(N,V,\{{\bf
r}_i\})\right]
\end{equation}
so that Eq.~(\ref{eq3.10}) can be viewed as an expansion of the
effective interaction between the CM in terms of (entropic) many-body
interactions, in a close analogy to expansion of the energy of atomic
or molecular systems in many-body interactions\cite{Carl90}.

\begin{figure}
\begin{center}
\centerline{\epsfig{figure=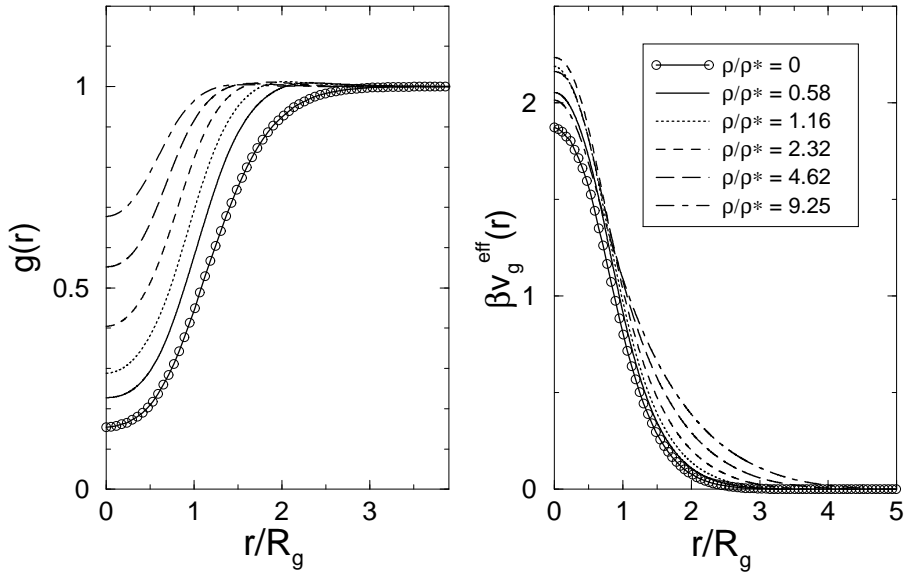,width=12cm}}
\caption{\label{fig:veffL500} The effective polymer pair potentials
$v_g^{eff}(r;\rho)$, derived at different densities $\rho/\rho*$ from
an Ornstein-Zernike inversion of the CM pair distribution functions
$g(r)$ of $L=500$ self avoiding walk polymer coils. (taken from
Ref.~\protect\cite{Bolh01a})}
\end{center}
\end{figure}

 But in practise following this path is extremely cumbersome, because
the number of coordinates and concomitant complexity of the
interactions $w^{(n)}$ grows rapidly with increasing order $n$.  To
circumvent this problem one could simply truncate the
expansion~(\ref{eq3.10}) at the pair level, but this would completely
ignore the many-body interactions.  Instead, we recently
proposed\cite{Loui00,Bolh01,Bolh01a} a coarse-graining method which
takes the many-body interactions into account in an average way.
First, at a given $\rho$, the $g(r)$ between the CM of a polymer was
generated by computer simulations.  Next, for each density, an
Ornstein-Zernike integral equation approach was used to invert the
$g(r)$ and generate the unique\cite{Hend74} effective pair potential
$v_g^{eff}(r;\rho)$ which exactly reproduces $g(r)$.  Explicit
examples of $v_g^{eff}(r;\rho)$ are shown in Fig.~\ref{fig:veffL500}
for different densities $\rho/\rho*$, where $\rho*=3/(4 \pi R_g^3)$
denotes the crossover from the so-called ``dilute'' to the
``semi-dilute'' regimes\cite{deGe79}.
For $\rho\rightarrow 0$, $v_g^{eff}(r;\rho) \rightarrow w^{(2)}(r)$
while the difference $\delta v_g(r;\rho)=v_g^{eff}(r;\rho)-w^{(2)}(r)$
grows with increasing density.

\begin{figure}
\begin{center}
\centerline{\epsfig{figure=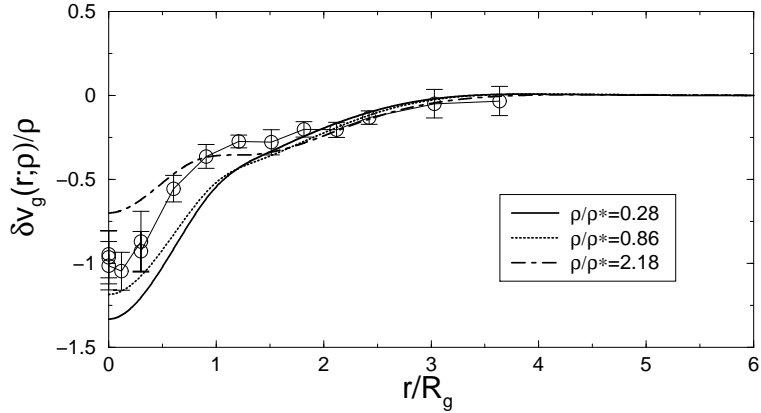,width=10cm}}
\caption{\label{fig:uex}Explicit simulations of
Eq.~(\protect\ref{eq3.5}) show that the three-body interaction
$w^{(3)}({\bf r}_i,{\bf r}_j,{\bf r}_k)$ is the dominant cause of the
density dependence of $v_g^{eff}(r;\rho) = w^{(2)}(r) + \delta
v_\rho(r;\rho)$ at lower densities. (taken from
ref.~\protect\cite{Bolh01a})}
\end{center}
\end{figure}

For the polymer system, the expansion~(\ref{eq3.5}) for the density
dependent correction to the effective pair potential, $\delta
v_g(r;\rho)$, can be explicitly calculated\cite{Bolh01a}.  The
results, plotted in Fig.~\ref{fig:uex}, show that despite the
existence of higher order interactions $w^{(n)}$ with $n > 3$ this
weak $w^{(3)}$ form performs remarkably well, demonstrating that for a
polymer solution, the three-body interaction is the dominant cause of
the density dependence in $v^{eff}_g(r;\rho)$ at least for $\rho/\rho*
\leq 1$.

\begin{figure}
\begin{center}
\centerline{\epsfig{figure=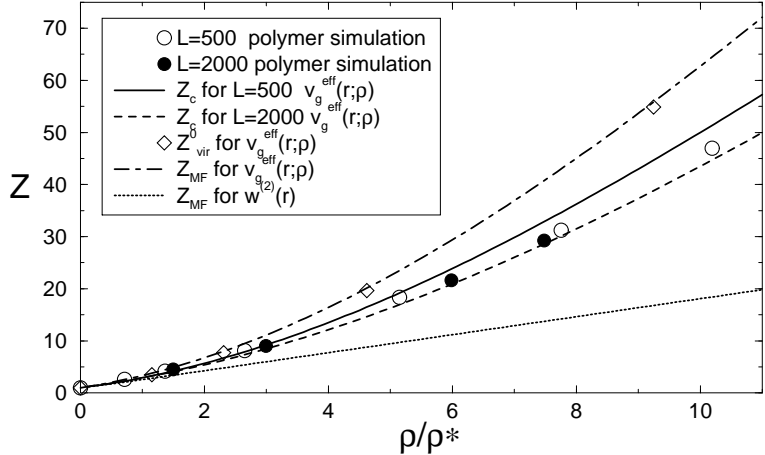,width=10cm}}
\caption{\label{fig:Z-linear} Equation of state of a polymer solution
calculated in several ways: Circular symbols are for direct
simulations self avoiding walk polymers; the two $Z_c$ follow from
Eq.~(\ref{eq2.1}); within the statistical errors of the method, they
agree with the direct simulations of the full polymer system.
$Z_{vir}^0$ was calculated by computer simulations with
$v_g^{eff}(r;\rho)$\protect\cite{Bolh01}, and agrees well with the
simpler $Z_{MF}$.  In contrast to $Z_c$, this is not a good estimate
for the true EOS.  Finally, when $Z_{MF}$ is calculated without any
density dependence of the pair potential, i.e.\ with $w^{(2)}(r)$
only, it strongly underestimates the EOS.  }
\end{center}
\end{figure}

Since $v_g^{eff}(r;\rho)$ was explicitly constructed to reproduce the
correct pair correlations, it follows that the true thermodynamics of
the full many-body system should be reproduced by using this potential
in the compressibility equation~(\ref{eq2.1}). The results shown in
Fig.~(\ref{fig:Z-linear}) confirm this: full polymer simulations of
the EOS $Z$ for $L=500$ and $L=2000$ self avoiding walk polymers on a
cubic lattice compare well with the EOS $Z_c$ calculated with the
appropriate $v_g^{eff}(r;\rho)$. The small residual differences at
large $\rho$ are most likely due to numerical difficulties in
performing accurate inversions at these high densities\cite{Bolh02}.

We now have a concrete example where the explicit construction of
$v_g^{eff}(r;\rho)$ leads to the full thermodynamics of a many-body
system within a purely pairwise description.  But what happens when
this density dependent pair potential is used within the virial
equation~(\ref{eq2.2})?  Fig.~\ref{fig:Z-linear} shows that the virial
equation without the explicit density derivative is accurately
approximated by the mean-field form $Z_{vir}^0 \approx Z_{MF}=1+ \rho
\hat{v}(k=0;\rho)$.  Just as was found for positive $\epsilon_1$ in
section~\ref{sec:2}, $Z_{vir}^0$ overestimates the true $Z$, and
adding the density-dependent correction (not shown) results in a
significantly larger overestimate.

These results provide a partial explanation for the thermodynamic
inconsistencies found in section~\ref{sec:2}. If $v(r;\rho)$ is
equivalent to $v_g^{eff}(r;\rho)$ then only the compressibility
equation has an unambiguous physical interpretation.  On the other
hand, one can also follow a route similar to that used to derive
$v_U^{eff}(r;\rho)$ to obtain a $v_{vir}^{eff}(r;\rho)$ which will
reproduce the true pressure of an underlying many-body system through
the virial equation\cite{Casa70,Hoef99,virial}. This potential will
not equal $v_g^{eff}(r;\rho)$, and so won't generate the correct pair
correlations needed in Eqs.~(\ref{eq2.2}) or~(\ref{eq2.9}). For an
underlying many-body system, no {\em single} pairwise density
dependent pair potential $v(r;\rho)$ exists which can, through either
the full density dependent~(\ref{eq2.2}) or the
density-independent~(\ref{eq2.9}) forms of the virial equation, yield
the true thermodynamics.

Nevertheless, Fig.~\ref{fig:Z-linear} shows that the relative
overestimate found using $v_g^{eff}(r;\rho)$ in the virial expression
is less than the relative underestimate found when all many-body terms
are ignored by only taking $w^{(2)}(r)$ into account.  Using
$v_g^{eff}(r;\rho)$ in the simple virial expression~(\ref{eq2.11}) is
therefore a better approximation than either ignoring density
dependence all together (as with $w^{(2)}(r)$), or using the full
density dependent form~(\ref{eq2.2}) of the virial equation.  In fact
for the dilute regime, $\rho/\rho* \leq 1$, where the polymers as soft
colloids coarse-graining technique is most useful, the absolute
differences between the virial and compressibility routes to the EOS
are quite small.  Differences in the structure generated by the two
effective potentials are also rather small\cite{Bolh02}.  On the other
had, for the semi-dilute regime ($\rho/\rho* > 1$), the polymer EOS
scales as $Z \propto \rho^{1/(3 \nu -1)} \approx \rho^{1.3}$, where
the Flory exponent $\nu \approx 0.59$\cite{deGe79}.  At these higher
densities the RPA approximation is excellent, which suggests that the
dominant density dependence of the FT of $v_g^{eff}(r;\rho)$ scales as
$\hat{v}_g^{eff}(0;\rho) \sim \alpha \rho^{0.3}$, with $\alpha$ a
density independent constant.  This implies that $Z_c \sim
(\alpha/2.3) \rho^{1.3}$, while $Z_{vir}^0 \approx Z_{MF} \sim
(\alpha/2) \rho^{1.3}$.  Using $v_g^{eff}(r;\rho)$ in the virial
equation at high densities results in the correct scaling exponent,
but a prefactor which is about $15\%$ too high.  In contrast, ignoring
the many-body terms altogether by using only $w^{(2)}(r)$ in either
the virial or the compressibility equations, results in $Z \sim \rho$,
which scales with the wrong exponent.

\subsection{Lessons for coarse-graining many-body systems}

The two previous subsections have shown that there is no unique way to
represent the effects of many-body interactions in effective pair
interactions.  This has important implications for several techniques
to derive such effective pair potentials.  For example, if an
ab-Initio or other higher level approach is used to generate the
structure of a many-body system, then the effective pair potential
that reproduces this structure will not be the same as the pair
potential that correctly describes the internal energy.  When the
many-body interactions are weak, these differences may not be that
important, but when they are strong, they may be significant.
However, as seen for the polymers as soft colloids approach, using
$v_g^{eff}(r;\rho)$ in the virial equation is still a better
approximation than completely neglecting all many-body interactions.
One might hope that the same is true for methods that derive effective
pair potentials from internal energies or from structure.

In general, the stronger the many-body interactions, the more likely
that extra care must be exercised when applying an effective pair
potential derived by one route (i.e.\ structure) to extract other physical
quantities (i.e.\ energy).  This has implications for work done on
systems with strong angular forces like water, where rather
complicated effective pair-potentials have be constructed to mimic
certain properties\cite{Male02}.

A more physically motivated way to average over many-body interactions
may be to derive effective pair interactions with a local density
dependence, since when the local instantaneous density of a liquid is
higher than the global average, one expects the relative strength of
the many-body interactions to be more important there, and vice versa
for lower local densities.  In fact, for a number of systems, one can
show explicitly that a local density dependence is equivalent to
many-body interactions\cite{Carl90,Pago00} (the same can be done for
internal degrees of freedom\cite{Sear00}).  Although this doesn't
imply that any many-body interaction can be consistently mapped onto a
local density dependence, for those cases where it can be done, a
completely self-consistent thermodynamics should exist based on these
effective potentials.

 Another obvious problem with a global density dependence arises when
one tries to treat inhomogeneous systems.  Even for Argon, the LJ pair
potential generates a surface tension that differs by up to to $19 \%$
compared to calculations that include explicit three-body
effects\cite{Bark93}. Again, a local density dependence would appear
the more natural way to coarse-grain.  Of course this opens up new
problems, such as how exactly does one define the local density
etc$\ldots$\footnote{An interesting recent example for an NVT
simulation can be found in\protect\cite{Alma01}.  However, the
original effective potential studied was of the $v_U^{eff}(r;\rho)$
form\protect\cite{Teje98}, while in~\protect\cite{Alma01} it is used
in a virial equation like Eq.~(\ref{eq2.2}).  The lack of
thermodynamic self-consistency found by these authors stems in part
from their use of a density dependent interaction without a careful
analysis of its origin.}

\section{Example 2: effective one-component Hamiltonians for 
two-component systems}
\label{sec:4}

Effective state dependent potentials naturally arise from tracing out
one component in a two-component mixture. For uncharged systems such
procedures were first carefully formulated by McMillan and Mayer in
their famous theory of solutions\cite{McMi45}.  As they, and many
other authors have stressed (see e.g.~\cite{Rowl84,Lekk92,Dijk98}),
this works most naturally in an ensemble where the component to be
traced out is treated grand-canonically.  For charged systems this is
no longer the case; instead, charge-neutrality makes the canonical
ensemble the natural choice.

As the following pedagogical examples will illustrate, whereas tracing
out procedures for uncharged systems are quite well understood, for
charged systems the waters are still muddied.

\subsection{Asakura Oosawa and related models}

\begin{figure}
\begin{center}
\centerline{\epsfig{figure=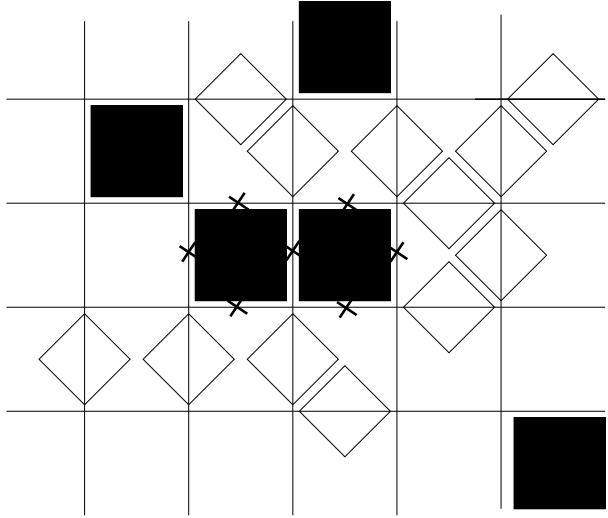,width=8cm}}
\caption{\label{fig:Lattice} A 2-dimensional lattice version of the AO
model. The big squares live on the $N$ sites, and the small ones live
on the $2 N$ links.  The big squares are hard, and cannot overlap each
other, and similarly the big and small squares exclude each other. The
small squares are ideal w.r.t.\ each other, and so multiple occupancy
of links not excluded by big particles is possible. The crosses on the
central big squares show which links they exclude from the small
particles. }
\end{center}
\end{figure}

A popular model for polymer-colloid mixtures consists of treating the
colloids as simple hard spheres (HS) of radius $R_{c}$ and the
polymers as penetrable sphere (PS) whose interaction with the colloids
is HS like, with a cross diameter $\sigma_{cp}$ of order $R_{c} +
R_g$, but whose interaction with other the polymers is ideal gas
like\cite{Asak58}.  Here I introduce a very simple and exactly
solvable lattice version of this Asakura Oosawa (AO) model, described
in Fig.~\ref{fig:Lattice}.

In the grand canonical ensemble the partition function for the mixture
becomes
\begin{equation}\label{eq4.1}
\Xi_{mix}(\mu_b,\mu_s,N) = \sum_{\{n_i\}} \sum_{\{m_i\}} \exp\left(
\mu_b \sum_i n_i + \mu_s \sum_i m_i \right)
\end{equation}
where $\mu_i$ denotes the chemical potential of each species, and the
sums are over all possible configurations $\{n_i\}$ and $\{m_i\}$ of
the big (``colloidal'') and small (``polymeric'') particles
respectively. Since the model is athermal, one can set $\beta = 1$; in
addition, the de Broglie wavelength is set to $\Lambda=1$ which
simplifies notation throughout this section.

For a given configuration of the big particles $\{ n_i \}$, there
will be $M(\{n_i\})$ links left for the small particles.  Their
partition function can then be calculated
\begin{equation}\label{eq4.2}
\Xi_{small}(\{n_i\})  =  \sum_{l = 0}^M \frac{M^lz_s^l}{l!} 
 = \exp[z_s M(\{n_i\})]
\end{equation}
where the fugacity $z_s=\exp[\beta \mu]$.  For a fixed configuration
$\{n_i\}$ of big particles, the number of free links is:
\begin{equation}\label{eq4.3}
           M(\{ n_i \}) = 2N - 4 \sum_i n_i + \sum_{<i\,j>} n_i \, n_j
\end{equation}
since there are two links for each site, each big particle excludes
$4$ links, but when two big particles touch, they only exclude $7$
links, because one is doubly excluded, as illustrated by the crosses
in Fig.~\ref{fig:Lattice}. This increase in the effective volume
available (and concomitant entropy) for the small particles when two
big ones touch is the origin of the osmotic depletion
effect\cite{Asak58}. A large enough fugacity of the small particles
can induce an entropy driven phase-separation of the big particles.

 With the explicit form~(\ref{eq4.3}) for the available ``free
volume'' $M(\{n_i\})$, the full partition function~(\ref{eq4.1})
simplifies to
\begin{equation}\label{eq4.4}
\hspace*{-1cm} \Xi_{mix}(\mu_b,\mu_s,N) = \exp \left[2 z_s N\right]
\sum_{\{n_i\}} \exp\left[(\beta \mu_b - 4 z_s) \sum_i n_i + z_s
\sum_{<i\,j>} n_i \, n_j \right].
\end{equation}
This illustrates the essence of the McMillan-Mayer tracing out
procedure: the sums over states in the partition function of the
original two-component system are rewritten as a single sum over the
states of an effective one-component system. There is no need to
invoke any Born-Oppenheimer type time-scale separation arguments; the
procedure might even be viewed as simply a mathematical trick. One
could just as well trace out the big particles.  But the particular
advantage of tracing out the small particles is that the effective
one-component system is simpler to treat, since it can be mapped onto
a binary lattice gas model with an effective chemical potential
$\mu_b^{eff} = \mu_b -4 z_s$ and an effective Hamiltonian that, in
turn, is decomposable as a pairwise sum over nearest neighbours with
an interaction strength $\beta \epsilon^{eff} = -z_s$.  By mapping
onto an Ising spin system in 2-D, this model, with its entropically
driven phase-separation transition, can be exactly solved, as shown
previously for a closely related model of non-additive hard
squares\cite{Fren92}.  Mathematically these models are a special cases
of the more general decorated Ising model introduced by
Widom\cite{Wido67}.

Keeping both species grand-canonical facilitates the mapping to an
Ising model, but it is just as easy to integrate out the small
particles in a semi-grand ensemble, i.e.\ fixing $N_b,z_s,N$,
resulting in an effective partition function of the form:
\begin{equation}\label{eq4.5}
Z_{mix}(N_b,z_s,N) = \exp\left[z_s(2N - 4N_b)\right] \sum_{\{n_i\}'}
\exp\left[z_s \sum_{<i\,j>} n_i \, n_j \right]
\end{equation}
where $\{n_i\}'$ denotes all possible ways of arranging $N_b$
particles on a lattice of $N$ sites.  Again, as long as $z_s$ is
fixed, $Z_{mix}$ can be interpreted as an effective one-component
system, interacting with an effective Hamiltonian of the form
\begin{equation}\label{eq4.5a}
\beta H^{eff}(N_b,z_s,N;\{n_i\}) = H_{bb} -z_s 2N + z_s 4 N_b - z_s
\sum_{<i\,j>} n_i \, n_j 
\end{equation}
with a configuration independent ``volume term'', and a pairwise
decomposable pair interaction.  $H_{bb}$ is the bare hard-core big-big
interaction. All the standard statistical mechanics for such
one-component systems can now be brought to bear to calculate
correlations and phase-behaviour.  The volume term contributes to
thermodynamic quantities, but not to the
phase-behaviour\cite{Dijk98,Liko01}.

The tracing out procedure can also be done in the canonical ensemble,
keeping both $N_b$ and $N_s$ fixed:
\begin{eqnarray}\label{eq4.6}
\hspace*{-2cm} Z_{mix}(N_b,N_s,N) &=& \sum_{\{n_i\}'}
\left( M(\{n_i\})\right)^{N_s} \nonumber \\
\hspace*{-2cm} & = & \sum_{\{n_i\}'} \exp\left[ N_s\log\left[2N - 4 N_b +
\sum_{<i\,j>} n_i \, n_j\right] \right]
\end{eqnarray}
But now the effective one-component system is no longer equivalent to
a system with a pair-decomposable effective Hamiltonian; at best it
can be rewritten as:
\begin{equation}\label{eq4.7}
\hspace*{-3cm} H^{eff}(N_b,N_s,N;\{n_i\}) = H_{bb} - N_s
\log\left[2N-4N_b \right] - N_s \log \left[1 +\frac{1}{2N-4N_b}
\sum_{ij} n_i \, n_j \right]
\end{equation}
In the limit of only two big particles the logarithms can be expanded
and pair term in the Hamiltonian reduces to the same form as found in
the same limit for the grand-canonical tracing out procedure.  But for
a larger number of big particles this ceases to be true; the
Hamiltonian can no longer be written as a sum over {\em independent}
many-body interactions of the form of Eqs.~(\ref{eq3.1})
or~(\ref{eq3.10}). The effect on $H^{eff}(N_b,N_s,N;\{n_i\})$ of
changing the number of pairs in a configuration $\{n_i\}$ by one
depends on the configuration of all other pairs in the system.  The
McMillan-Mayer mathematical tracing out procedure does not lead to
such a useful simplification in the canonical ensemble as it does in
the semi-grand ensemble.

Similar manipulations can be performed for off-lattice two-component
systems such as the original AO model\cite{Asak58}.  Integrating out
the $N_p$ PS polymeric particles for a fixed configuration $\{{\bf
r}_i\}$ of $N_c$ colloidal HS results in effective Hamiltonians of the
form:
\begin{eqnarray}
H^{eff}(N_c,z_p,V;\{{\bf r}_i\}) &=& H_{cc} + \Omega(N_c,z_p,V;\{{\bf
r}_i\}) \label{eq4.20a}\\ H^{eff}(N_c,N_p,V;\{{\bf r}_i\}) &=& H_{cc} +
F(N_c,N_p,V;\{{\bf r}_i\}) \label{eq4.20b} 
\end{eqnarray}
for the semi grand and canonical ensembles respectively. $H_{cc}$ is
the bare colloid Hamiltonian. $\Omega(N_c,z_p,V;\{{\bf r}_i\})$ is the
grand-potential and $F(N_c,N_p,V;\{{\bf r}_i\})$ is the Helmholtz free
energy of an {\em inhomogeneous} system of PS particles in the
external field of the colloids.  The free energy form of these
effective Hamiltonians holds for a general two-component
system\cite{Dijk98}.  Integrating over all possible configurations
$\{{\bf r}_i\}$ of the colloidal particles leads to the full
two-component partition function. For the AO model these Hamiltonians
take a particularly simple form since
\begin{equation}\label{eq4.21a}
\Omega(N_c,z_p,V;\{{\bf r}_i\}) = -z_p V^{free}(N_c,V;\{{\bf r}_i\})
\end{equation}
is the grand potential of an ideal gas at in the accessible free
volume $V^{free}(N_c,V;\{{\bf r}_i\})$ and  similarly
\begin{equation}\label{eq4.21b}
\hspace*{-1cm} F(N_c,N_p,V;\{{\bf r}_i\}) = N_p \log[N_p] - N_p -
N_p\log\left[V^{free}(N_c,V;\{{\bf r}_i\})\right].
\end{equation} is the Helmholtz free energy of an ideal gas of $N_p$
particles in the accessible free volume.

The calculation of $V^{free}(N_c,V;\{{\bf r}_i\})$, the direct
analogue of $M(\{n_i\})$ in the AO lattice models, simplifies
dramatically if the size-ratio $R_g/R_c \leq 0.1547$. There are then
no triplet or higher order overlaps of the exclusion zones, and the
accessible free volume can be written as
\begin{equation}\label{eq4.22}
V^{free}(N_c,V;\{{\bf r}_i\}) = V - N_c V_1 + \sum_{i<j} V_2(r_{ij})
\end{equation}
where $V_1= \frac43 \pi (\sigma_{cp})^3$ is the volume excluded by
each colloidal particle, and $V_2(r)$ has the standard AO
form\cite{Asak58}, depending only on the relative separation $r$ of
two particles.  (Note the correspondence with Eq.~(\ref{eq4.3}) for
the lattice model, where $V=2N$, $V_1=4$ and the sum over $V_2(r)$ is
replaced by a nearest neighbour lattice sum). The effective
Hamiltonians of Eqs.~(\ref{eq4.20a}) and~(\ref{eq4.20b}) simplify to
\begin{eqnarray}
\hspace*{-1.5cm} H^{eff}(N_c,z_p,V;\{{\bf r}_i\}) &=& H_{cc}  - z_p V + z_p N_c V_1 -
z_p\sum_{i<j} V_2(r) \label{eq4.23a}\\
\hspace*{-1.5cm}H^{eff}(N_c,N_p,V;\{{\bf r}_i\}) &=& H_{cc} + N_p
\log[\frac{N_p}{V-N_c V_1}] - N_p \nonumber \\ &-& N_p \log\left[1 +
\frac{1}{V-N_c V_1}\sum_{i<j} z_p V_2(r) \right]. \label{eq4.23b}
\end{eqnarray}
Just as was found for the lattice model, the canonical tracing out
procedure leads to a Hamiltonian that cannot be written as a sum over
independent interactions of the form of Eqs.~(\ref{eq3.1})
or~(\ref{eq3.10})\footnote{The small differences between the volume
terms in the lattice and off-lattice versions of the AO model stem
from kinetic energy terms that are ignored in the former case.}.  In
contrast, semi-grand $H^{eff}(N_c,z_p,V,\{{\bf r}_i\})$ takes on the
more useful form of a ``volume term'' plus a sum over an effective
pair potential.  The interpretation of the latter will now be treated
in more detail.

\setcounter{footnote}{0}
While volume terms in Eq~(\ref{eq4.23a}) are important to make contact
with the full two-component system, as worked out very clearly and in
great detail by Dijkstra van Roij and Evans\cite{Dijk00}, they do not
contribute to the phase-behaviour of the system.  For $R_g/R_c <
0.1547$ the simplified pairwise Hamiltonian
\begin{equation}\label{eq4.24}
H^{eff}(N_c,z_p,V) = H_{cc} + \sum_{i<j} v^{eff}_z(r;z_p)
\end{equation}
with $v^{eff}_z(r;z_p) = - z_p V_2(r)$, completely determines the
phase-behaviour and correlations of the colloidal system.  An entirely
self-consistent one-component {\em osmotic} thermodynamics can be
defined based on the Hamiltonian~(\ref{eq4.24}).  This means, for
example, that the compressibility equation~(\ref{eq2.1}) generates the
same {\em osmotic} EOS as the virial equation:
\begin{equation}\label{eq4.25}
Z_{vir} =\frac{\Pi(N_c,z_p,V)}{\rho_c} = 1 - \frac{2\pi}{3} \beta  \rho_c
\int_0^\infty r^3 \frac{\partial v^{eff}_z(r;z_p)}{\partial r}
g_{cc}(r) dr,
\end{equation}
where $\Pi(N_c,z_p,V)$ is the osmotic pressure\footnote{Note that
adding the explicit volume term $\partial F_0/\partial V$ to the
compressibility and virial equations would generate the thermodynamics
of the full two-component system.}.  Since $z_p$ is a fixed {\em
external parameter}, $v^{eff}_z(r;z_p)$ is in fact not state-dependent
within this one-component picture.  Thus, the McMillan-Mayer tracing
out procedure results in a system that can be {\em exactly} mapped
onto a classical one-component fluid with a state-independent pair
potential\footnote{For larger size-ratios, the mapping results in
three-body and higher order interactions, leading to similar problems
to those discussed in section~\ref{sec:3} when one tries to derive an
effective pair potential. Also, interactions between the small
particles will lead to effective three-body and higher terms.}.  One
must simply keep in mind that the resultant thermodynamics quantities
are {\em osmotic}, and defined w.r.t.\ a reservoir containing PS
particles only.  Similarly, since the McMillan-Mayer procedure is
essentially a mathematical trick to facilitate calculating the sums in
the partition function, very little information can a-priori be
extracted about the effective {\em dynamics} of the one-component
system.  For that, further physically motivated arguments and
approximations are needed.

Now suppose one were working in the canonical ensemble, and naively
used the Asakura-Oosawa depletion potential $v^{eff}_z(r;z_p)$, as has
sometimes been done in the literature. Then $z_p = z_p(\rho_c,\rho_p)$
would no longer be an external parameter but would instead depend on
the state of the system. For example, for small size-ratios a good
approximation is $z_p \approx \rho_p/(1 - \rho_c
V_1)$\cite{Dijk98,Dijk00}.  One might be tempted to define an
effective one-component thermodynamics by fixing $\rho_p$, and
treating it as an external parameter, as might be natural in an
experiment. The potential then takes on the $\rho_c$ dependent form
$v_\rho^{eff}(r;\rho_c;\rho_p)$. If this potential is taken as given,
without enquiring as to its origins, as was done in
section~\ref{sec:2}, then the compressibility and virial equations
take forms similar to Eqs.~(\ref{eq2.1}) and~(\ref{eq2.2})
respectively, since $v_\rho^{eff}$ depends on $\rho_c$.  One might
expect that the ensuing thermodynamics would again be that of the
osmotic system, but now {\em neither} the derived virial nor the
derived compressibility equation is correct.  For the present case of
a fixed $\rho_p$, the compressibility equation~(\ref{eq2.1}) using
$v_\rho(r;\rho_p,\rho_c)$ does not result in the correct osmotic
pressure because $z_p$ varies when $\rho_c$ is changed.  On the other
hand, using $v_\rho^{eff}(r;\rho_c,\rho_p)$ in the simpler virial
equation~(\ref{eq4.25}), i.e.\ one without a density derivative, does
generate the correct osmotic EOS, as follows from the following
arguments: The thermodynamic properties are independent of ensemble,
so, for a given state point $(N_c,N_p,V)$, the osmotic pressure is the
same as in the semi-grand ensemble at a state point $(N_c,z_p,N)$ such
that $<N_p>_{zp,Nc,V}=N_p$.  There the potential
$v_\rho^{eff}(r;\rho_p,\rho_c)=v_z^{eff}(r;z_p)$ generates the correct
pair correlations $g_{cc}(r)$, and also the correct virial pressure
$\Pi(N_c,N_p,V)$ through Eq.~(\ref{eq4.25}).  But the apparent
relevance of this osmotic virial equation within the canonical
ensemble is deceptive -- it only follows because it can be derived in
the semi-grand ensemble, and used at a state point where
$v_\rho^{eff}(r;\rho_c,\rho_p) = v_z^{eff}(r;z_p)$.

 Of course the lack of consistency between virial and compressibility
routes should not be surprising, since a careful tracing out procedure
demonstrates that the effective Hamiltonian~(\ref{eq4.23b}) for the
canonical ensemble {\em cannot} be decomposed into a sum over
independent pair potentials.  Therefore, using the Asakura-Oosawa pair
potential in this ensemble is not rigorously justified, except for
instances where parallels with the semi-grand ensemble can be
made\footnote{At this point I should point out that in the limit
$N_c=2$, both the canonical and the grand-canonical ensembles do
result in the same pair potential, as can be seen by expanding the
$\log$ in Eq.~(\ref{eq4.23b}). A similar conclusion holds for higher
order interactions, should they be relevant.  However, I stress that
great care must be used when these low density effective interactions
are applied at finite colloid densities in the canonical ensemble.}.

In summary then: for a fixed $z_p$, a completely self-consistent
one-component thermodynamics can be derived in the semi-grand ensemble
for an AO system.  But within the canonical ensemble, a McMillan-Mayer
style tracing out procedure leads to an effective Hamiltonian that
cannot easily be written as a sum over pair and higher order
interaction terms; the ensuing one-component system does not have a
simple interpretation as an effective liquid.  Moreover, if the AO
depletion potential is naively applied in the canonical ensemble for
fixed $\rho_p$, then the density-dependence of the pair potential
again leads to an apparent lack of consistency between different
routes to osmotic thermodynamics.  In contrast to the case studied in
section~\ref{sec:3}, where the effective density dependent potential
$v_g^{eff}(r;\rho)$ that arises from tracing out three-body
interactions generates the correct thermodynamics only through the
compressibility equation, here only the virial equation~(\ref{eq4.25})
should be used.

\subsection{Debye H\"{u}ckel model}

The effective interactions and resulting phase-behaviour of
charge-stabilised colloidal suspensions have been the subject of much
recent debate\cite{Roij97,Bell00}.  In contrast to uncharged mixtures,
global charge-neutrality implies that the canonical ensemble is the
natural choice in which to integrate out the  co and counter
ions,  to arrive at an effective one-component colloidal
picture. And this explains in part why the problem is so difficult,
since, as was shown in the previous subsection, tracing out one
component in the canonical ensemble does not necessarily lead to an
obvious description in terms of independent (many-body) interactions.
In addition, direct computer simulations of the full mixture are
greatly complicated by the long-range nature of the Coulomb
interactions and the large length scale differences between a typical
colloidal particle, and the co- and counter-ions.

Rather than attempting yet another tracing out procedure, this section
has a much more modest goal, namely to illustrate pitfalls that arise
from a naive application of a very simple textbook density dependent
potential of the Debye-H\"{u}ckel screened Yukawa form
\begin{equation}\label{eq4.40}
\beta v_{DH}(r;\rho) = \frac{Z^2}{r} \exp\left[- \kappa(\rho) r\right].
\end{equation}
Here $Z$ is the charge of the colloidal particle, and
$\kappa(\rho)=\sqrt{4 \pi Z \rho}$ is the screening parameter in the
absence of salt.  The Bjerrum length $\lambda_B = \beta e^2/\epsilon$,
with $e$ the elemental charge and $\epsilon$ the dielectric constant,
has been set to $1$,to simplify the notation.  Since $\kappa$ depends
on the overall density (through charge neutrality), it should come as
no surprise that a simple application of the compressibility
equation~(\ref{eq2.1}) and the virial equation~(\ref{eq2.2}) do not
generate the same thermodynamics.  Since Eq.~(\ref{eq4.40}) is an
integrable potential,
\begin{equation}\label{eq4.40a}
\beta \hat{v}(k=0;\rho)= \frac{4 \pi Z^2}{\kappa^2} = \frac{Z}{\rho},
\end{equation}
 its thermodynamic behaviour resembles that of a mean field
fluid\cite{Loui01a} for large $\rho$ or small effective $Z$, where the
RPA closure should be quite accurate.  Thus the two routes lead to:
\begin{equation}\label{eq4.41}
Z_c^{RPA} = 1 + Z,
\end{equation}
which can be interpreted as the ideal EOS of the colloids and $Z$
counterions, and
\begin{equation}\label{eq4.42}
\hspace*{-2cm} Z_{vir} = (1 + \frac12 Z) - \frac{ 2 }{3} \beta \pi
\rho \int r^2 \left[h(r)(r \frac{\partial v_{DH}(r)}{\partial r}) -
g(r) 3 \rho \frac{\partial v_{DH}(r)}{\partial \rho} \right].
\end{equation} 
 Even the leading term in the virial equation differs from the
compressibility equation.  Since both the $r$ and $\rho$ derivatives
of $v_{DH}(r;\rho)$ are always negative, the second two terms of
$Z_{vir}$ both reduce its value w.r.t.\ the leading $Z_{MF}=(1 +
\frac12 Z)$ term\footnote{At least in the regime where the RPA is a
reasonable approximation. Note that the virial equation can be exactly
solved in the RPA approximation\protect\cite{unpublished}}, increasing
the difference between the two routes even further.  The present
discrepancy originates not in the lack of consistency of the closure,
but rather in the naive application of a density dependent pair
potential.  A more careful analysis of the underlying two-component
colloid $+$ counterion system shows that volume terms must also be
taken into account\cite{Roij97}, but these don't bring the two routes
any closer together.  The only way to know which (if any) of the two
routes is the more reliable would be to derive them from a careful
analysis of the full tracing out procedure\footnote{One example where
Eq.~(\protect\ref{eq2.2}) appears to be better than forms without the
density derivative can be inferred from the discussion of the
Debye-H\"{u}ckel model by Belloni\protect\cite{Bell00}.  By adding the
explicit volume term to his application of the Debye-H\"{u}ckel
potential in Eq.~(\protect\ref{eq2.2}), the virial pressure closely
resembles an approximate form derived from a two-component picture.}

The arguments above may have repercussions for the modelling of
liquid metals by effective density-dependent pair potentials.  There,
the Ascarelli-Harrison form of the virial equation~(\ref{eq2.2}) has
often been applied, but as other examples in this paper show, it is
not yet clear whether this equation is reliable.  In fact, a recent
careful derivation of an effective one-component virial equation
within within a two-component electron ion picture by Chihara {\em et
al.}\cite{Chih01} does not include an explicit density derivative.

Finally, it appears that effective potentials in charged systems will
always be density dependent due to charge-neutrality.  But this same
charge neutrality implies that a two component system can be viewed
effectively as a one-component one.  For example, all the partial
structure factors have the same $k\rightarrow 0$ limit, up to trivial
prefactors, i.e.\ they are not independent.  The density dependence in
charged systems most likely has a different character from that found
in uncharged systems.

\section{Conclusions}\label{sec:conclusions}

The overall conclusion of this paper is simply that

\begin{quotation} \noindent {\bf an effective
density dependent pair potential ${\bf v(r;\rho)}$ cannot be properly
interpreted without reference to the coarse-graining procedure by
which it was derived}.  

\end{quotation}
 
\noindent This was shown by a number of explicit
examples.  The only systems where an effective pair potential could be
rigorously interpreted as part of an effective Hamiltonian were the AO
model for $R_g/R_c \leq 0.1547$ and its lattice analogue.  Other
potentials like $v_g^{eff}(r;\rho)$ cannot be formally interpreted as
part of an effective Hamiltonian.  Instead, they should be viewed as
mathematical devices to calculate desired properties from a particular
equation\cite{Bark69,Casa70,Rowl84,Hoef99}.

Some of the more detailed points are repeated below:

\begin{itemize}

\item A pair potential that depends on global density, $v(r;\rho)$,
does not lead to the same thermodynamics in the canonical and
grand-canonical ensembles.

\item The status of the Ascarelli-Harrison\cite{Asca69} form of the
virial equation~(\ref{eq2.2}), with an added $\partial
v(r;\rho)\partial \rho$ term, is suspect.  For uncharged systems and
many-body systems, counter examples where it does not apply and even
makes thermodynamic inconsistencies worse, can easily be constructed.
Whether it is valid for liquid metals or other charged systems remains
to be proved.

\item There is no unique way to represent the effect of many-body
interactions as density dependent pair
interactions\cite{Bark69,Casa70,Rowl84,Hoef99}.  The potential that
correctly reproduces the structure, $v_g^{eff}(r;\rho)$, will not
generate the right internal energy when used in the energy equation
etc...  At best, one can pick a $v^{eff}(r;\rho)$ that performs well
for the particular physical properties one is interested in. For
example, the parameters commonly used in the LJ potential to model
Argon can be viewed as a compromise between those that correctly
reproduce the energy, the virial pressure, and the
structure\cite{Hoef99}.

\item The previous statement implies a link between problems with the
{\em transferability} of an effective pair potential, and problems
with its {\em representability}.  For example, if many-body
interactions generate a large relative density dependence in
$v^{eff}(r;\rho)$, then the potential derived by a given route at
$\rho_1$ will differ significantly from the one derived at $\rho_2$
(transferability problems).  This large density dependence also
implies important differences between $v_U^{eff}(r;\rho)$ and
$v^{eff}_g(r;\rho)$ (representability problems).

\item A McMillan-Mayer type tracing out procedure for a two-component
system in the canonical ensemble does not lead to an effective
one-component Hamiltonian that be written as a sum of independent
interactions.

\item The Debye-H\"{u}ckel potential, like all such density dependent pair
potentials, leads to thermodynamic inconsistencies when it is naively
used in an effective  one-component picture.  At present it is not yet
clear what the correct route to thermodynamics is.

\item Obvious problems arise when potentials like $v(r;\rho)$ are used
in inhomogeneous systems.  There it would make more sense to
coarse-grain to potentials that depend on a measure of the local
density.  In some cases, this can be shown to be equivalent to a
many-body interaction approach.

\end{itemize}

I finish with the question: when do these results matter?  The picture
is not as bleak as I might have painted it in the discussions above.
Indeed, in the case of Argon, the differences between
$v_g^{eff}(r;\rho)$, and $ v_U^{eff}(r;\rho)$ are not so large, so
that a single compromise effective pair potentials works admirably
well for the liquid phase. In addition, many examples can be found for
soft-matter systems where ignoring the many-body forces altogether, --
simply using $w^{(2)}(r)$ in the expansion of an $H^{eff}$ -- works
quite well.  This is true for many systems described by depletion
potentials\cite{Meij94,Dijk98,Dijk99}, or for the effective
interactions derived for star-polymers\cite{Liko98} or
dendrimers\cite{Liko01a}.  And in other cases, such as the polymers as
soft colloids example of section~\ref{sec:3}, using
$v_g^{eff}(r;\rho)$ in the simple virial equation may still be a
better approximation than ignoring many-body effects altogether.
Nevertheless, the particular coarse-graining procedure used to derive
$v^{eff}(r;\rho)$ must always be kept in mind -- the naive consumer of
effective pair potentials should beware.

\section*{Acknowledgements}
I acknowledge support from the Isaac Newton Trust, Cambridge, and
thank Neil Ashcroft, Peter Bolhuis, Bob Evans, Daan Frenkel, Martin
van der Hoef, Gerhard Kahl, Hartmut L\"{o}wen, Paul Madden, Richard
Sear, Vincent Krakoviack, and Patrick Warren for many useful
discussions.  But most of all, I would like to thank Jean-Pierre
Hansen, who has been closely involved in this work, and who has been a
great inspiration to me, both as a scientist and as a person.  It is
with great pleasure that I dedicate this paper to the honour of his
60th birthday.


\begin{thebibliography}{0}
\bibitem{virial} An effective potential $v_{vir}^{eff}(r;\rho)$ can be
defined that generates the same pressure as the full three-body virial
equation when used in the standard two-body virial equation without a
density dependent contribution (see e.g.\protect\cite{Hoef99} for an
example, where $\delta v_{vir}(r;\rho)=3 \delta v_U(r;\rho)$).  But since
this effective potential is density dependent, one might also use the
density dependent form of the virial equation, which would generate a
different effective potential.
\end{thebibliography}
\end{document}